\documentclass[11pt,twoside,A4]{article} 
\usepackage{times,fancyhdr}
\usepackage[dvips]{graphicx}
\usepackage{latexsym} 
\usepackage[affil-it]{authblk}
\usepackage{amsmath}
\usepackage{amssymb}
\usepackage{setspace}
\usepackage{hyperref}

\usepackage[top=4cm, bottom=4cm, left=3.5cm, right=3.5cm]{geometry}

\def\beq{\begin{equation}}
\def\eeq{\end{equation}}
\def\beqn{\begin{eqnarray}}
\def\eeqn{\end{eqnarray}}

\begin{document}
 
\title{Theory and Phenomenology of Spacetime Defects}
\author{Sabine Hossenfelder\thanks{hossi@nordita.org}} 
\affil{\small Nordita\\
KTH Royal Institute of Technology and Stockholm University\\
Roslagstullsbacken 23, SE-106 91 Stockholm, Sweden}
\date{}
\maketitle
\begin{abstract}
Whether or not space-time is fundamentally
discrete is of central importance for the development of the
theory of quantum gravity. If the fundamental description of space-time is discrete,
typically represented in terms of a graph or network, then the
apparent smoothness of geometry on large scales should be
imperfect -- it should have defects. Here, we review a model
for space-time defects and summarize the constraints on the prevalence of
these defects that can be derived from observation.
\end{abstract}

\section{Introduction}

A theory of quantum gravity is necessary to describe the quantum behavior
of space and time and to understand what happens in strong gravitational fields,
when curvature reaches the Planckian regime.
Finding this missing theory of quantum gravity is one of the big open problems in theoretical physics
today, and it concerns the most fundamental ingredients of our existing
theories: Space-time and its curvature, the arena in which physics happens.

But general relativity still stands apart from the quantum field theories of the standard model as
a classical theory. There exists to date no known way to consistently couple a classical theory
to a quantum theory, and neither do we know how to quantize gravity. While several theoretical approaches
are being pursued with success, this success has so far been exclusively on the side of mathematical
consistency, and the connection of these approaches to reality is still unclear. 

The problem how to resolve the tension between quantum field theory and
general relativity is more than an aesthetic unease. This tension signals that our
understanding of nature is incomplete, but it also offers an opportunity to improve
our theories. The missing theory of quantum gravity has the potential to revolutionize our understanding of space, time
and matter.

Progress on the theory of quantum gravity however has been slow. The problem
has been known since more than 80 years now. Since then we gained a great many insights
about the nature of the problem but the big breakthrough has left us waiting.
Next to the technical difficulties, the reason for the slow progress is
lack of experimental guidance. The possibility that quantum gravitational phenomena
might be observable has not been paid much attention to till the late 90s, and
even now the awareness that this possibility exists is slow to sink into the
minds of the community. However, without making contact to observation, no theory
of quantum gravity can ever be accepted as a valid description of nature.

In the absence of a fully-fledged theory, 
this search for observable consequences proceeds by the development
of phenomenological models. Such models parameterize properties 
that the theory of quantum gravity could have with the purpose of allowing
to experimentally test or at least constrain the presence of these
properties. This in turn guides the 
development of the theory.  General reviews of phenomenological models for
quantum gravity can be found in \cite{Hossenfelder:2010zj, AmelinoCamelia:2008qg}.
In this contribution to the {\sc AHEP} special issue on `Experimental Tests of Quantum Gravity
and Exotic Quantum Field Theory Effects' we will discuss a possible phenomenological consequence of quantum gravity
that has so far received very little attention -- the existence of space-time defects.

In many approaches to quantum gravity -- such as causal sets, spinfoams, causal dynamical
triangulation, loop quantum gravity, and emergent and induced gravity scenarios based on
condensed matter analogies -- space-time is fundamentally
discrete and the smooth background geometry that we see only emerges as
an approximation at low energies and large distances \cite{Loll:1998aj}. In this case, one expects that
the apparently smooth background geometry is imperfect and has defects, just
because perfection would require additional explanation.  

In the following, we will review the recently proposed model
for space-time defects \cite{own1,own2} and summarize the constraints on the prevalence of
these defects that can be derived from observation. We will then discuss which steps can be
taken to improve the model so that the constraints touch on the interesting parameter range.

\section{Space-time Defects}
 
Whether or not space-time is fundamentally
discrete is a question of central importance for the development of the
theory. But this discreteness typically makes itself noticeable at the Planck
length, which is hard if not impossible to access experimentally. Thus, instead of
searching for direct evidence for the Planck scale discreteness, we here propose
 to look for imperfections in this discreteness. Such imperfections in
space-time will cause deviations from general relativity, since general relativity 
rests on the assumption that space-time is a manifold and locally smooth and
differentiable. 
Because general relativity is an extremely well-tested theory even smallest 
deviations can become noticeable, making space-time defects a promising
phenomenological consequence to search for. Looking for space-time defects as evidence for the presence of
a discrete geometry is akin to looking for specks of dirt as evidence for the
presence of a window. 

That such defects should exist is a model-independent expectation for
all approaches to quantum gravity in which geometry has a fundamentally 
discrete structure. But the prevalence, distribution, and properties of the defects will depend
on the details of the underlying fundamental theory. This way, the phenomenological
model can bridge the gap between theory and experiment. In the following we will aim
to parameterize the consequences of space-time defects so that contact can be
made to the underlying theory if the relevant parameters can be extracted.

In contrast to defects
in condensed matter systems, space-time defects are not only localized in
space but also in time. They do not have worldlines but are space-time
events. Space-time defects can come in
two different versions, local defects and nonlocal defects. The general case
would be a hybrid of both, but treating these two types separately is helpful 
to develop the theory. We will
first discuss the nonlocal defects.

\subsection{Nonlocal Space-time Defects}

One of the reasons it is expected that quantum gravity necessitates nonlocality
is that the notion of the Planck
length as a minimal length implies that it is meaningless to distinguish
points below this distance -- these points are not local (or not points, depending
on your perspective). But
besides this, during
the last decades it has also become increasingly clear that a resolution
of the black hole information loss problem requires some type of nonlocality \cite{Giddings:2009ae}.
Meanwhile, a completely different line of investigation has led to the conclusion
that requiring the Planck energy to be an observer-independent component
of a four-momentum, necessitates to give up absolute locality.  Instead one might
have to settle on a weaker locality requirement, which has been called `relative locality' \cite{AmelinoCamelia:2011bm}. 

While Planck-scale nonlocality has received quite some attention, for
example in the well-studied models for quantum field theory with a minimal
length \cite{Hossenfelder:2012jw}, here, we will focus on a feature whose
phenomenology has so far received very little attention: Macroscopic 
nonlocality, where macroscopic means much larger than the Planck length.

Macroscopic nonlocality can be expected to arise in approaches towards
quantum gravity in which space-time is only to good approximation a smooth
manifold but fundamentally a network (graph) consisting of nodes and links which can
carry additional charges or degrees of freedom. The reason is that the emergence of a manifold from
the network will not be perfect, but it will have defects. And since our
macroscopic notion of distance only emerges with the space-time and the
metric that we define on it, there is no reason why these defects should
respect the emergent macroscopic locality. 

This has been demonstrated explicitly by
Markopoulou and Smolin in \cite{Markopoulou:2007ha} for the case of spin
networks. Their argument can be briefly summarized as follows. 

States of the spin-network describe spatial slices of space-time
and they change in time by a set of allowed evolution moves, which are local
according to the locality of the network.
Certain spin network states, called `weave states', match to good precision (up to Planck scale corrections), 
slowly varying classical spatial metrics.
The structure of nodes and links of the network carries information about area and volumes 
and thus the geometry that the network fundamentally describes. 
It can then be shown that it is possible to act with a large number of
local evolution moves on a state without nonlocal links and by this 
create a state that contains macroscopically nonlocal links. This is
possible without changing the classical state that it approximates, thus
the existence of the classical approximation alone cannot be used to
rule out these nonlocal links. 

\begin{figure}[th]
\includegraphics[width=10cm]{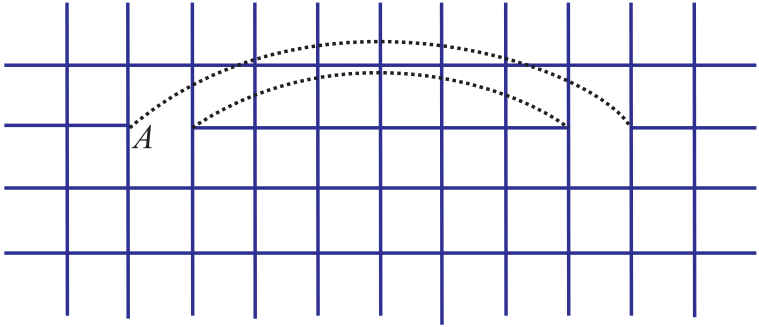}

\hspace*{2cm}\caption{Schematic picture for nonlocal links on a regular lattice. The lattice represents space-time. The link, which represents a defect in the regularity of the background, is long according to the distance measure of the background. Note that all nodes remain 4-valent. If space-time is fundamentally discrete, similar defects should be all around us.  \label{fig1}}
\end{figure}

A nonlocal link is, intuitively, a link in the network
that does not respect the emergent macroscopic locality. More strictly it can be identified
by the number of nodes on the shortest closed loop that it is part of.
For the nonlocal link this will be a large number of nodes, while all the local links have a small 
number of nodes (depending on the valence of the network). See Figure \ref{fig1} for illustration.

In the example by Markopoulou and Smolin, the states with nonlocal links 
are still allowed solutions and thus valid semi-classical space-times, but do not respect the macroscopic locality. 
Because of simple combinatorics, one
finds that there are in fact many more nonlocal states than local states.
This means the locality of the
state that we live in today is not perfect; nonlocal links should
be all around us. This situation was aptly dubbed `disordered locality' in \cite{Markopoulou:2007ha}. 

The argument by Markopoulou and Smolin is an explicit
example one can have in mind. But the expectation for the existence
of nonlocal defects is more general than that, it arises because
perfection requires additional explanations or selection criteria that we do not have.

Now such macroscopic locality might strike one as something to be avoided,
since we do not seem to experience it, but this is a question of experimental 
constraints. To really understand the implications of such nonlocality one first 
needs to develop a phenomenological model that parameterizes
the effects, and then contrast them with data.  Such a model was
developed in \cite{own1}. This model does not start with an underlying
discrete structure, but instead deals with the defects in the local structure
as deviations from the smooth background geometry of general relativity. 

The central assumption for the model \cite{own1} is that Lorentz-invariance is
preserved on the average  since violations of Lorentz-invariance are strongly disfavored by the
data.
Lorentz-invariance, maybe not so surprisingly, proves to be very restrictive on the type of nonlocality
that is allowed. The distribution of nonlocal defects in this model is assumed to be given
by the only presently known Lorentz-invariant discrete distribution on
Minkowski space, which is defined by the Poisson-process described in 
\cite{Dowker:2003hb, Bombelli:2006nm}.
With this distribution, the probability of finding  N points in a space-time volume $V$ is
\beqn
P_{\rm N}(V) = \frac{(\beta V)^{\rm N} \exp(-\beta V)}{{\rm N}!} \quad, \label{pois}
\eeqn
where $\beta = L^4$ is a constant space-time density and $L$ a parameter of
dimension length.

A particle that encounters a nonlocal defect will experience a translation
in space-time. The translation vector is parameterized in a probability
distribution which besides $L$ introduces a parameter of dimension
mass, $\Lambda$, and a parameter of dimension length $\alpha$. 
$\Lambda$ and $\alpha$ (both real-valued and positive) quantify the translation, $y_\nu(\alpha, \Lambda)$
that the particle experiences at the non-local defect
\beqn
p_\nu y^\nu = \alpha \Lambda \quad, \quad y_\nu y^\nu = \pm \alpha^2 \quad, \label{sign}
\eeqn
where the choice of sign determines whether the translation is timelike or
spacelike. The interpretation of $\Lambda$ is roughly speaking that a
particle will be translated a distance of about $\alpha$ in the restframe in which
its energy is about $\Lambda$. 

The translation that the particle experiences when it encounters the nonlocal
defect is then given by a probability distribution $P_{\rm NL}(\alpha, \Lambda)$ 
over the endpoints. By construction, this is all entirely Lorentz-invariant. The density
$\beta$ together with the distribution $P_{\rm NL}(\alpha, \Lambda)$ determines 
the phenomenology of the model. One can further simplify this situation by 
approximating the probability distribution by a Gaussian with mean values $\langle \alpha \rangle, \langle \Lambda \rangle$
and variances $\Delta \alpha$, $\Delta \Lambda$. This leaves one with 5 parameters.
These can be further reduced by assuming that there is only one new length scale
$L \sim \langle \alpha \rangle$, and that the width of the distributions is comparable
to the mean value $\langle \alpha \rangle \sim \Delta \alpha$ and $\langle \Lambda \rangle \sim \Delta \Lambda$.
One is then left with two parameters, one length scale and one mass scale, that can be
constrained by experiment quite simply. While this might not be the most general
case, it allows one to get a first impression on what amount of nonlocality is
compatible with observation.

A massless particle that encounters a nonlocal defect will be deviated
from the lightcone and on the average travel either faster or slower than the normal
speed of light (depending on the choice of sign in (\ref{sign})). It is possible to restrict translations to be timelike
and velocities to be subluminal to
avoid causality problems because such a restriction does not violate
Lorentz-invariance. The repeated scattering on nonlocal defects
creates a small effective mass for the photon, that is the mass that
a particle of the photon's energy would have on the average
trajectory that contains nonlocal links. This is reflected in the
average speed of the particle which, in the presence of nonlocal defects, can deviate from the speed of
light because the translations that the particle experiences when it
encounters a nonlocal defects may be spacelike or timelike rather
than lightlike. 

It is important to note that the distribution
of translation vectors in this model is not an independent property of space-time but
depends on the wave-function of the incident particle, in the simplest case its
momentum vector, in the general case the average momentum vector and the
spatial width of the wave-function (at the moment it encounters the defect). It is this dependence
on the incident particle 
that allows one to construct a normalizable and Lorentz-invariant
distribution. While the full Lorentz-group is non-compact, using measurable
properties of the incident particle as reference prevents the
need to introduce a Lorentz-invariance violating cutoff while
still maintaining observer-independence. 

The central feature that distinguishes this model from other
random-walk like models for propagation in a quantum space-time
is that the probability of the particle being affected by the quantum
properties, here the space-time defects, depends on the (Lorentz-
invariant) world-volume
that is swept out by the particle's wavefunction. Thus, the larger
the position uncertainty of the particle and the longer its propagation
time, the more likely the particle is to be affected by the space-time
defects. This generically means that particles of long wavelength
are better suited to find phenomenological consequences than highly
energetic ones, in contrast to the phenomenology that arises for
example within deformations or violations of Lorentz-invariance \cite{AmelinoCamelia:2008qg,AmelinoCamelia:2010pd}.

With the use of this model for nonlocal space-time defects, constraints on the 
density of defects and the parameters of the model can then be derived from
various observables. 

\begin{figure}[ht]

\hspace*{-0.5cm}\includegraphics[width=16cm]{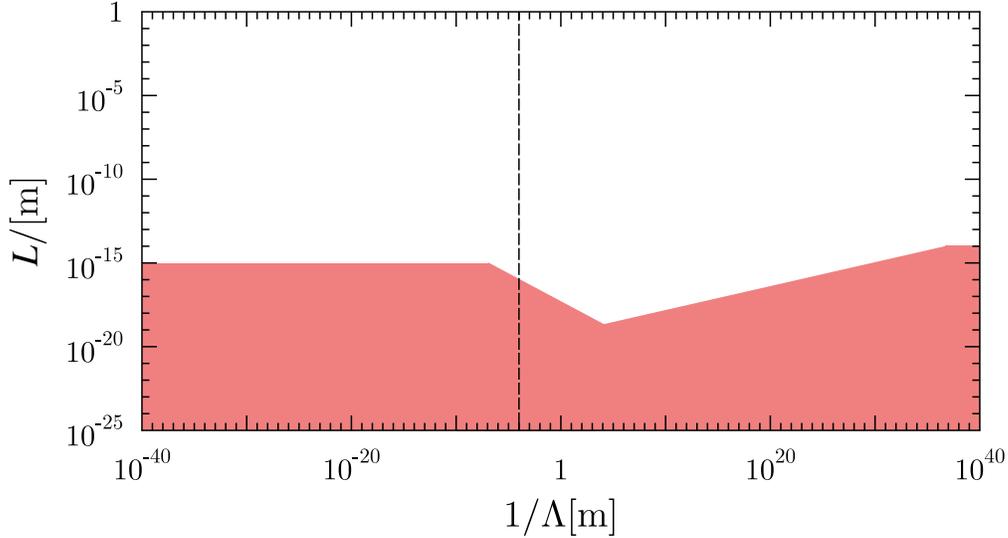}

\caption{Summary of constraints on nonlocal space-time defects, from \protect{\cite{own1}}. The red shaded region is excluded. The dashed (black) line
indicates the value of the cosmological constant.  \label{combined}}
\end{figure}

For example, we have good evidence that 
protons of ultra-high energy which give rise to cosmic ray showers 
originate in active galactic nuclei. The protons have a finite mean-free
path when traveling through the cosmic microwave background ({\sc CMB})
because they can scatter on the {\sc CMB} photons and produce pions. In
the presence of nonlocal defects, the protons' mean free path can
increase because the particles effectively do not travel the
full distance. If the mean free path increases substantially, this would
be in conflict with observation and can thus be used to derive bounds
on the density of defects. Other constraints come from the blurring of
interference rings in images of distant quasars, and from the close monitoring of single photons in a
cavity. For details and references please refer to \cite{own1}. 
The constraints on nonlocal defects can be visually summarized
in Figure \ref{combined}. Roughly speaking it can be concluded that the density
of nonlocal defects has to be less than one in a space-time volume
of fm$^4$. 

In the cosmological context, a natural scale is given
by the length scale associated to the measured value of the cosmological
constant, which is about $1/10$~mm. The constraints on $L$ are
presently about 10 orders of magnitude below this interesting parameter
range. One can expect however that the existing data is actually sensitive
to larger values of $L$. It's just that the model in its present
form cannot be used to reliably analyze much of the existing cosmological
data because it does not take into background curvature. It is clearly
desirable to generalize the model to at least a Friedmann-Robertson-Walker
background to be able to analyze cosmological data for evidence of
space-time defects.

\subsection{Local Space-time Defects}

A model for local defects can be developed based on a similar
approach as that for the nonlocal defects \cite{own2}. Much like
the nonlocal defects cause a statistically distributed translation in
position space, local defects cause a statistically distributed 
translation in momentum space. In other words, the local defect
stochastically changes the momentum of the incident particle, thereby violating
momentum conservation. Since
the defect transfers a distribution of momenta, it consequently has
a finite size. The finite size of the defect, together with the 
preservation of locality make the local defects much easier to treat
and incorporating them into a quantum field theoretical framework
is relatively straight-forward.

The
distribution of local defects is again assumed to be given by the
Poisson sprinkling (\ref{pois}). The momentum non-conservation
is then parameterized in the length scale $L$ of the distribution and a mass scale $M$.
These could a priori be different from the parameters relevant for local
defects. 

The coupling of quantum fields to the local defects is incorporated
by adding a term to the gauge covariant derivative $\partial + eA \to
\partial + e A + g \partial P$. Here $A$ is some gauge field with
coupling constant $e$, $g$ is a coupling constant for the space-time
defects and $P$ is essentially the Fourier transform of the distribution
of the momentum that is stochastically transmitted by the defect.
This then allows one to calculate the probability for different
scattering processes in an $S$-matrix expansion as usual. Importantly, this
particular coupling to the defects has the effect that an on-shell
particle that scatters on a defect is necessarily off-shell after
scattering. 

For a massless particle with energy $E$ in 1+1 dimensions, the assumption that
the momentum has a Gaussian distribution over the model parameters
leads to a space-time defect that also has a Gaussian distribution in
lightcone coordinates
\beqn
P(x^+, x^-) &=& \exp\left(\frac{(x^+)^2}{(2 \sigma^+)^2} + \frac{(x^-)^2}{(2 \sigma^-)^2}\right) 
\frac{e^{{\rm i} (\langle k_+ \rangle x^+ + \langle k_- \rangle x^-)}}{2 \pi \sqrt{\sigma^+ \sigma^-}}
\quad,
\eeqn
with widths
\beqn
\sigma^+ = \frac{\sqrt{2} E}{\Delta M^2} \quad, \quad \sigma^- = \frac{\sqrt{2}}{E} \quad, \label{sigmas}
\eeqn
where $\Delta M^2$ is the variance of the distribution of the parameter $M^2$. 
One sees that the typical space-time patch covered by the defect is
\beqn
\sigma^+ \sigma^- = 2 \left(\Delta M^2 \right)^{-1} \quad.
\eeqn
The defect has a Lorentz-invariant volume independent of $E$, though it will deform 
under boosts that red- or blueshift $E$ as one sees from Eqs. (\ref{sigmas}). Care must
be taken in higher dimensions to properly normalize the momentum distribution. As
with the nonlocal defects, the normalization can be achieved using the same
method that is commonly used in the evaluation of scattering amplitudes, by
taking into account that in reality we strictly speaking never deal with plane
waves. The finite spatial extension of the incident particle's wavefunction
serves to regularize the distribution. 

Massive particles can be treated similar to the massless ones. Again, it
is of central relevance that the momentum distribution of the defect is
a function of the properties of the incident particle. For massive particles,
the distribution can be assigned most easily in the restframe of the
particle, and in that restframe it will have an especially simple form.
Thus, while the distribution is not Lorentz-invariant in the sense that its expression
changes under arbitrary Lorentz-transformations, observer-independence
is maintained because all observers can use the incident particle's momentum
as a reference and obtain the same result.

With this setup, constraints can be derived from processes normally
forbidden in the standard model, which are now allowed. The most
important bounds come from long-lived particles that travel long
distances and are the following:
\begin{enumerate}
\item {\bf Photon decay:} After scattering on a defect, a photon 
is off-shell and subsequently decays into a fermion pair. This effect is similar to
pair production in the presence of an atomic nucleus in standard quantum
electrodynamics ({\sc QED}).
This process results in
a finite photon lifetime, and leads to excess electron-positron pairs. 
\item {\bf  Photon mass:} The presence of space-time
defects makes a contribution to the photon propagator and creates a
small photon mass. (Gauge invariance is violated.) 
\item {\bf Vacuum Cherenkov radiation:} An electron can emit
a (real) photon after scattering on a defect. This is similar
to QED Bremsstrahlung.
\end{enumerate}

\begin{figure}[ht]
\hspace*{0.5cm}\includegraphics[width=14cm]{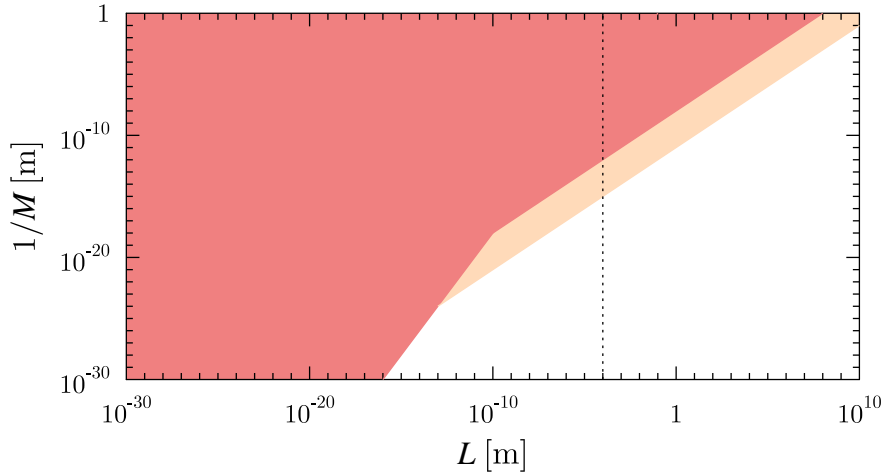} 

\caption{Summary of constraints on local space-time defects, from \protect{\cite{own2}}. The red (dark) shaded region is excluded. The peachy (light) shaded region indicates
a stronger constraint from photon decay with the ad-hoc assumption that the typical distance between 
defects increases with the cosmological scale factor.  \label{fig3}}
\end{figure}

 The constraints from these effects can be summarized in Figure \ref{fig3}. 
Again, note that the existing bounds on $L$ are several orders of 
magnitude, but not too far, below the interesting parameter range.
(Making $1/M$ smaller than shown in the plot means it becomes
comparable to the Planck length $\sim 10^{-35}$~m and then
it doesn't make sense any more to speak of defects.) It would
thus be highly desirable to improve the model to tighten the bounds.

Not much
work has been done on local space-time defects prior to \cite{own2}, except for the
model proposed in \cite{Schreck:2012pf}. The model in  \cite{Schreck:2012pf}  
differs from the one discussed here in three important ways. 

First, in  \cite{Schreck:2012pf} 
the interaction with the defect is mediated by scalar field. Second, the treatment in
 \cite{Schreck:2012pf} necessitates the introduction of a cut-off in the
momentum-space integration which breaks Lorentz-invariance and
defeats the point of using a Lorentz-invariant distribution of defects
to begin with. Such a cut-off
is unnecessary in the model discussed here where the regulator is essentially the
spatial width of the wave-packet. Third, and most important, the coupling to the defect
in  \cite{Schreck:2012pf} is different. The approach in \cite{own2} started from the
assumption that the defects originate in a distortion of space-time
regularity and make themselves noticeable as a modification in
the covariant derivative. This leads to a specific structure of the
coupling terms, which was not used in  \cite{Schreck:2012pf}. 

In summary it can be said
that space-time defects are pretty much unexplored territory where
not much previous work has been done. That makes the topic
very exciting as it harbors a potential for breakthrough. 

\section{Discussion}

The preliminary
work \cite{own1,own2} tested the potential
of detecting space-time discreteness by the occurrence of defects in the background's regularity. The
models used in this preliminary work can deliver only 
rough estimates. They are suitable for these estimates, but are theoretically unsatisfactory. Since the estimates
show that it seems possible to reach the
interesting parameter ranges experimentally, a further investigation and improvement of
the theory and phenomenology of space-time defects is desirable. In the following we will
propose some steps into this direction.

First, the models proposed in \cite{own1,own2} are for flat 3+1 dimensional 
Minkowski space. Since the best constraints on the presence of local and
nonlocal defects come from particles that have traveled long
times and distances, one could
derive better constraints when background curvature in
general,
and an expanding Friedmann-Robertson-Walker ({\sc FRW}) metric
in particular, can be taken into account. This would then allow one to use cosmological
data, eg from the cosmic microwave background, to constrain the density of 
defects.

First steps towards a cosmology with nonlocal defects have been taken in references \cite{PrescodWeinstein:2009wq} and 
\cite{Caravelli:2012wy} based on
the quantum graphity model developed by Markopoulou et al \cite{Konopka:2008hp}. 
It was assumed in \cite{PrescodWeinstein:2009wq} that the
nonlocal connections lie within a timelike slice that is assumed to be identical to the
cosmological time. This of course violates Lorentz-invariance, but in a time-dependent
background this can be expected. However, this specific violation of Lorentz-invariance is very strong and 
artificial: There is really no reason why the
time-evolution of the network should be identical to the cosmological time, which is
only an approximation based on the assumption of a homogeneous matter distribution anyway. 
More realistically, one would expect both slicings to differ, so that the links would still
have a spread in the time-like coordinate. This would alter phenomenological 
consequences. The model for defects discussed here provides a good basis to
study this phenomenology.

Second, the model with
nonlocal defects also so far only operates on a kinematical
level and a full dynamical description in terms of a quantum
field theoretical treatment is missing. It would be desirable to development a 
quantum field theoretical model for this case, and thus also be able to
combine both local and nonlocal defects. One way to address this
point would be 
to use the dual nature of the local and nonlocal defects and to
make mathematically precise the idea that nonlocal defects
act like local defects, just in position-space rather than in
momentum space. This would allow one to express them as
operators on the particles' Hilbert space and facilitate their incorporation 
into quantum field theory.

Third, while the search for defects as a model-independent expectation for
space-time discreteness is interesting in its own right, contact to theoretical
approaches to quantum gravity would serve a better identification of the
parameters. The density of defects might for example be possible to
extract in approaches that display a phase-transition from a pre-geometrical
phase to an approximately smooth geometry. In this case the density of
remaining defects should depend on the properties of the phase transition.

\section{Conclusion}

If space-time is fundamentally of non-geometric origin, then the smooth background
geometry of general relativity should have defects. The consequences of these defects 
can be parameterized and described in phenomenological models, which allow one
to put constraints on the density of the defects and the strength of their effects. These
models are in their infancy and much remains to be done, but they harbor the possibility
that a targeted search for space-time defects will allow us to find evidence for
quantum gravity.

\end{document}